%% file: main.tex
\documentclass[12pt]{article}
\usepackage{graphicx}
\usepackage{fullpage}
\usepackage{float}
\usepackage{amsmath}
\usepackage{array}
\usepackage{graphicx}
\usepackage{subcaption}
\usepackage[numbers]{natbib}
\usepackage{url}

\begin{document}

\title{Boolean queries are all you need?}
\author{ 
  Charles L. A. Clarke \and
  Mark D. Smucker
}
\date{
University of Waterloo, Canada}

\maketitle

\begin{abstract}
We equipped an LLM-based search agent with access to a Boolean retrieval engine to search the MS MARCO V2.1 deduped segment collection used by the TREC 2024 RAG track.
Over a standard track subset of 86~topics~---~and operating under a budget of 100~model calls/topic~---~the agent achieved an NDCG@10 of 0.6863, which would place it above many dense, sparse, and learned-sparse first-stage retrievers.
Ranking is based solely on the density of corpus substrings matching a query~---~with no requirement for supervised learning, global statistics, or term weights.
Formally, the query language expresses a strict subset of the regular languages, with a document's score based on the number and length of matches it contains.
Although the results are more exploratory than definitive~---~because they are based on a single test collection that was publicly available during model training~---~they suggest that simple pattern matching may be sufficient for agentic search.
\end{abstract}

\section{Introduction}

Long before the emergence of AI-powered natural-language search and keyword-focused Web search, Boolean queries provided the primary interface to many operational search engines~\cite{sfw83}.
While Boolean retrieval was effective in the hands of librarians and other trained technicians~---~allowing them to specify precise requirements for retrieved results~---~it proved difficult for untrained people to successfully formulate queries.
As early as 1982, efforts were already underway to automatically reformulate natural language queries into Boolean queries, simplifying interaction~\cite{salton82}.
By the mid-1990s, Boolean retrieval was fading away.
Keyword-based sparse vector retrieval was well established in academic contexts, BM25 was gaining a reputation for excellent effectiveness, and the earliest keyword-based Web search engines were widely available~\cite{okapi,metacrawler,witten1994managing}.
Even so, as late as 1995, manually constructed Boolean queries could still outperform the best automatic systems at TREC~\cite{clarke1995shortest,harman1995trec4overview}.
A few years later, using interactive searching and judging with Boolean queries, trained searchers independently created judgments yielding system rankings consistent with official TREC-6 rankings, translating human intuition into exact-match queries to find relevant documents~\cite{isj98,voorhees98}.
Nonetheless, given that untrained humans often struggle to formulate effective Boolean queries, by the 21st century, they were largely abandoned for general-purpose search.

\input{goth}
\input{quebec}

Large language models face no such struggle.
When appropriately prompted, LLMs can reformulate natural language into reasonable-seeming Boolean queries~\cite{adam2024literature,10.1145/3673791.3698432,10.1145/3539618.3591703,wang2025reassessinglargelanguagemodel}.
Figure~\ref{fig:goth} provides an example from the experiments in this paper.
This example can be compared to the human-generated query in Figure~\ref{fig:quebec}.
In both cases, the Boolean query includes terms not found in the original topics.
In both cases, the searcher was encouraged to use their internal knowledge as a guide to query expansion.

As a test of LLM-driven Boolean retrieval, we implemented a simple search agent (Vole) based on the interactive searching and judging (ISJ) procedure used by \citet{isj98} to independently create judgments for TREC-6 topics.
For retrieval, we adopt a newer implementation of their string-matching algorithm, which treats the corpus as a single contiguous string and efficiently identifies shortest substring matches to a Boolean expression. 
Documents are ranked based on the length and number of matches they contain, avoiding the need for global statistics and term weights (Section~\ref{sec:ssr}).

The Vole search agent acts as little more than a runner between a search engine and an LLM (Section~\ref{sec:vole}).
After providing instructions to the LLM, the Vole carries queries from the LLM to the search engine and brings results back, managing the interaction.
The interaction is limited to 100~model calls/topic. On each call, the model may choose one of three actions: 1) run a new query, receiving a 200-word snippet centered around the shortest substring in the top document; 2) ask for a snippet from the next document for a previous query; 3) ask to see a full document.
As documents and snippets are seen by the model, it returns relevance judgments, along with reasons to support the judgments.
In addition to the limit of 100~model calls, the Vole will halt the interaction after the model has made 50~distinct relevance judgments, even if the model-call limit has not been reached.

The instructions encourage the model to find as many highly-relevant documents as possible.
Since Vole searches by exploring, rather than ranking, we re-rank the discovered documents for comparison with other methods, using model-judged relevance as a primary key and discovery order as a secondary key.
We report experiments on a standard subset of TREC 2024 RAG Track topics, using gpt-5.5 as the LLM (Section~\ref{sec:rag24}), comparing the results to a range of official track submissions from the first author's team~\cite{monster}, with performance ranging from a typical BM25 baseline to one of the top-3 runs (Section~\ref{sec:results}).
Since the Vole digs up documents it judges highly relevant but that are unjudged, we use UMBRELA~\cite{upadhyay2024largescale,upadhyay2024umbrela} to fill holes by assessing these previously unjudged documents, rather than relying on Vole's own assessments (Section~\ref{sec:rag24}).

\section{Related Work}

This paper is only one of several recent papers suggesting that traditional IR methods may be sufficient to support retrieval augmented generation and agentic search.
As early as 2023, \citet{doostmohammadi-etal-2023-surface} demonstrated that, in retrieval-augmented language modeling, retrieval gains are better explained by surface-form overlap than by dense semantic similarity, and that BM25-style surface retrieval can reduce language-model perplexity.
Also in a RAG context, \citet{OldIR} report that ancient unsupervised methods such as Markov Random Fields (MRF) and relevance-model retrieval outperform both BM25 and dense retrieval.

Multiple papers appearing in May 2026 focused specifically on unsupervised retrieval methods for agentic search.
\citet{yang2026superintelligentretrievalagentfrontier} propose a Superintelligent Retrieval Agent (SIRA), which uses an LLM to construct a single corpus-discriminative BM25 query rather than relying on multi-round exploratory search.
In contrast with our work, SIRA concentrates the model's effort into constructing one ``ultimate'' BM25 query, while Vole gives the model an exploratory loop for revising Boolean queries as evidence is observed.
Also using an exploratory loop, \citet{hsu2026rethinkingagenticsearchpiserini} find that BM25 can support effective agentic search when paired with a model that can retrieve, browse, and read documents interactively.

Most closely connected with our work,
\citet{sen2026grepneedagentharnesses} compare grep-style lexical search with vector retrieval inside an agentic harness, finding that simple exact-match tools can be highly competitive when agents control the search process.
\citet{li2026semanticsimilarityrethinkingretrieval} also propose direct corpus interaction, in which agents search raw corpora using exact-match terminal-style tools rather than fixed top-$k$ retrieval interfaces.
Like us, neither approach depends on vectors, learned weights, or global statistics, operating solely on surface-level textual matching.
Unlike them, we support indexing for efficiency, including immediate transactional update, and ranking based on the length and density of matches.

\section{Shortest Substring Ranking}
\label{sec:ssr}

We treat the corpus as a single contiguous string and solve boolean queries under {\em minimal interval semantics}~\cite{bv16,bv18,ssr,clarke1995shortest}.
Each solution to a query is a substring of the corpus satisfying the Boolean query that itself has no proper substring satisfying the query (a ``shortest substring'').
By using an extended inverted index, we can enumerate these substrings in $O(m \log n)$ time, where $n$ is the number of occurrences of the most frequent term, and $m$ is the number of solutions, i.e., shortest substrings in the corpus that satisfy the query~\cite{buttcher2010information,clarke2025annotative,ssr}.

For ranking, we score a document on the basis of the shortest substrings it contains, making two assumptions: 1) The shorter the substring, the more likely the containing document is relevant. 2) The more substrings contained in a document, the more likely it is relevant.
As we enumerate the shortest substrings, we compute a score for each document containing at least one substring, maintaining a current top-$k$ list.

To compute a score for a document, let \({\cal S}\) be the set of shortest substrings contained in a document, where each substring is expressed as an interval \((p_i,q_i)\), where the endpoints of the interval indicate positions in the corpus string.
For each \((p_i,q_i) \in {\cal S}\), \(p_i\) is the start position of the substring and \(q_i\) is the end position of the substring.
The shortest substring ranking (SSR) score of the document is then
\begin{equation}
  \operatorname{score}({\cal S})
    =
    \sum_{i=1}^{|{\cal S}|}
    \frac{1}{C + q_i - p_i}.
\end{equation}
\(C > 0\) is a smoothing constant, with default value $C = 42$ used in our experiments.

Each substring contributes inversely with its length, so shorter substrings contribute more to the document score than longer substrings.
Since the score is based solely on local properties of the document, no global statistics are required, and the ranking method is trivially parallelizable.
Since the languages expressible by our Boolean query language are a strict subset of the regular languages, shortest substring ranking can be viewed as a limited form of ranked regular expression retrieval.
Potentially, the same method could be applied to rank documents based on general regular expression matches.

For indexing and shortest substring ranking we use the Cottontail search engine\footnote{\url{https://github.com/claclark/Cottontail}}.
Cottontail implements a generalization of inverted indexes called ``annotative indexing,'' in which both content and metadata are represented as interval annotations over a common address space~\cite{clarke2025annotative}.
The same mechanism can support ordinary ranked retrieval, including BM25/WAND, structural queries over JSON objects, and sparse-vector annotations with a single unified indexing structure.
Unlike a conventional static retrieval index, Cottontail also supports dynamic annotation and transactional update, allowing new derived structure to be added to existing content and made visible immediately after commit while preserving ACID properties for concurrent readers and writers.
Thus, Cottontail is not merely a postings-list implementation for Boolean queries, but can provide a retrieval-native substrate for agent memory, where raw text, structure, annotations, and derived model judgments can coexist, be queried, and be updated as the agent works.

\section{Search Agent}
\label{sec:vole}

For the experiments in this paper, we implemented a simple search agent, the Vole\footnote{\url{https://github.com/claclark/TheVole}}.
The Vole search agent consists of three components: a remote model, a set of detailed
instructions, and a small interaction runner.
These components are supported by a specialized SSR server\footnote{\url{https://github.com/claclark/Cottontail/blob/main/apps/ssr-server.cc}} that uses the Cottontail C++ library to implement a dedicated search interface.
For our experiments, the agent is not given full access to Cottontail, and is limited to executing Boolean queries and accessing document contents.

The model (\texttt{gpt-5.5} for our experiments) is prompted with instructions, a topic description, and an accumulated interaction transcript.
The instructions describe the query language and the core concepts of shortest substring ranking, along with the JSON protocol for communication with the runner, and through it, the server.
At each step, the model must return exactly one JSON action: issue a new query, request the next result for an existing query, or request a full document.
The instructions ask the model to judge each newly observed document on a 0--3 relevance scale and to steer searches toward grade-3 evidence.

The runner is deliberately simple.
It maintains the interaction transcript, validates the model's JSON decisions, sends search actions to the retrieval service, and records both the model decisions and search responses.
The interaction transcript is maintained append-only to fit model-side caching requirements, minimizing costs and allowing the model to see the full interaction at each step.
The runner also enforces two stopping conditions: at most 50 unique document judgments per topic and at most 100 model calls, where the limit on model calls is primarily set to avoid run-away interactions in which the model is unable to formulate queries that return previously unjudged documents.
When a new query surfaces a duplicate already-judged document, the runner skips it.
The model is shown skipped document identifiers and their ranks but does not spend judgment budget on them again.

The runner communicates with the SSR search server using a simple newline-delimited JSON protocol over a local TCP socket.
For each model action, the runner sends one request to the server, such as a Boolean query, a request for the next result from a previous query, or a request for the full text of a document.
The server returns a JSON object containing either a ranked result with a document identifier and 200-word snippet, an exhausted-result marker, or the requested document text.
These server responses are appended to the transcript that is shown to the model on the next step.
Fuller request and timing information is retained in separate interaction and audit logs.

The SSR server itself is a lightweight JSON-line service that exposes shortest-substring ranking over one or more Cottontail collections.
While it was designed for use in this experiment, it is general enough to be usable by other search agents for other purposes.
A client sends a Boolean query, and the server evaluates it against a pre-configured definition of ``document'', ranks matching documents by the SSR substring score, and returns results one at a time with document identifiers, snippets, and timing information.
It maintains query state so clients can request additional results for a query and retrieve full documents.

\input{results}

\section{Task and Dataset}
\label{sec:rag24}

We base our experiments on the TREC RAG Track 2024 retrieval task, which is a standard adhoc task: Given a topic, return a ranked list of documents according to the probability ranking principle.
The track used the MS MARCO V2.1 deduped segment collection\footnote{
\url{https://trec-rag.github.io/annoucements/2024-corpus-finalization/}} as the corpus\footnote{
Since the collection is segmented, segments are ``documents'' for retrieval purposes.
For TREC 2024, the original MS MARCO V2.1 collection was segmented with a ``sliding window size of 10 sentences and a stride of 5 sentences''
to make the corpus ``more manageable for users and baselines.''
For SSR, the unit of retrieval is defined at query time, which for these experiments was always a pre-defined segment.
However, this unit of retrieval could be a dynamically defined window of words, or with the addition of annotation for sentences, a dynamically defined window of sentences.}.
Before starting these experiments, we already had the collection indexed with Cottontail as a single shard, used for some of the first author's TREC 2024 experimental runs~\cite{monster}.
This supports a direct comparison with those runs, since the corpus representation is held fixed while only query processing and ranking change.

Of the 301~track topics, a subset of 86~topics have (partial) manual judgments, while the remainder have only LLM-based judgments from the UMBRELA judgment tool~\cite{upadhyay2024umbrela}.
Official results were reported for both the full 301-topic set and the 86-topic subset~\cite{upadhyay2024largescale}.
For reasons of parsimony, we base our experiments on the 86-topic subset, but use UMBRELA judgments in all cases.

\section{Results}
\label{sec:results}

Figure~\ref{fig:results} compares results from the Vole with a selection of official track submissions from the first-author's team\cite{monster}.
For comparison purposes, a Vole interaction is collapsed into a TREC run by sorting the documents judged by Vole first by the agent's own assigned relevance grade and then by discovery order.
The external qrels are not used to construct this ranking; they are used only afterward by \texttt{trec\_eval}.
When the agent surfaced unjudged documents, we used UMBRELA to fill these holes with the same remote model (gpt-4o) used for the official judgments.
UMBRELA judged 438 of these previously unjudged documents to be at least highly relevant.

The agent achieved an NDCG@10 of 0.6863.
In the context of the 77~official runs submitted to the track, this performance would place it above many dense, sparse, and learned-sparse first-stage retrievers, but below many rankers that themselves employed LLM-based re-ranking.
For example, \texttt{uwc1} was a top-3 run for the task.
It is based on a reciprocal rank fusion (RRF)~\cite{rrf} of 15 different ranking stacks, including traditional BM25 and dense rankers, with and without LLM-based query expansion.
The resulting fused ranking (\texttt{uwc0}) was then pointwise and pairwise reranked using LLM-based relevance scoring methods~\cite{pref,ictir23} to produce \texttt{uwc1}.
In terms of effort, the creation of \texttt{uwc1} required an average of 194 model calls per topic for expansion and reranking vs.\ less than 74 for the Vole. 
Other groups used various LLM-based re-ranking methods~\cite{pradeep2023rankvicuna,pradeep2023rankzephyr,sun2023rankgpt} at TREC 2024.
If we are using LLMs for re-ranking, model calls (and tokens) become important measures of cost.

When we checked efficiency in the logs, we were a little surprised by query latency.
In our past experience, a typical human-written query would run in under 100ms.
Only 15\% of Vole queries ran at that speed and 59\% required more than a second.
One query required well over a minute.
We traced the issue back to an unrecognized problem with Cottontail's phrase search optimization, now fixed.
Humans, or at least the handful who have written queries for SSR ranking, tend to use common collocations (``united states''), while in the slower-running queries, the LLM was using disjunctions of uncommon phrases containing high-frequency terms (``make this issue known''), which were then repeatedly re-solved by the engine, instead of being memoized.

\section{As They May Search}

We have come full circle.
Up until the 1980s, operational search engines depended on Boolean retrieval.
While Boolean queries allowed trained operators to precisely specify information needs, they proved difficult for untrained humans to employ effectively.
Thus, the history of information retrieval is~---~at least in part~---~a history of increasingly effective methods for searching with keyword and natural language queries, culminating in sophisticated semantic matching methods, such as dense dual-encoder retrievers like DPR and ANCE~\cite{karpukhin2020dense,xiong2021ance}, and late-interaction models such as ColBERT~\cite{khattab2020colbert}.
But while these methods account for human limitations in query formulation, large language models face no such limitations.
\citet{sen2026grepneedagentharnesses} and \citet{li2026semanticsimilarityrethinkingretrieval} ask if grep is all that search agents need.
Perhaps they need even less.

\section{Limitations}
We report experiments for a single task on a single test collection that was available publicly during model training.
Thus, the results should be viewed more as exploratory than definitive.
While the core of Cottontail was largely written by the first author banging on keys one at a time, the dedicated SSR server and the Vole were largely vibe-coded with Codex.
We applied sanity checking, reviewed data and logs, ran manual tests, and asked the coding agent a lot of questions, but we did not review every line of code.
This paper was written with AI assistance, but all of the organization and much of the original wording come from the authors.
The shard used for the experiments includes field-oriented TF-IDF annotations, and the experiment could easily be repeated with BM25F.
We have not done this yet.

\section*{Acknowledgments}

This work would not be possible without the many years of insights and contributions from Gordon Cormack.
We also thank Andrew Parry for his helpful comments.

\bibliographystyle{plainnat}
\bibliography{main}

\end{document}

%% file: goth.tex
\begin{figure}[t]
\begin{verbatim}
(^ "goth rock" (+ society social political issues themes addressed))
(^ "goth rock"
   (+ darkness depravity alienation despair death)
   (+ lyrics themes))
(^ goth
  "post-punk"
  (+ politics political "social commentary" society)
  (+ lyrics themes))
\end{verbatim}
\vspace*{-\baselineskip}
\caption{Example LLM-generated queries for the TREC 2024 RAG Track topic ``what society issues did goth rock address'' (\#2024-152259).}
\label{fig:goth}     
\end{figure}

%% file: quebec.tex
\begin{figure}[t]
\begin{verbatim}

(^
  (+
    (^ quebec (+ separatism separatist separation separate))
    (^ quebec (+ sovereignty sovereign independent independence))
    "jacques parizeau"
    "louise beaudoin"
    "parti quebecois")
  (+ economic economically economics poll opinion referendum majority))
\end{verbatim}
\vspace*{-\baselineskip}
\caption{Example human-generated query for the TREC-4 adhoc topic ``What are the prospects of the Quebec separatists achieving independence from the rest of Canada?'' (\#207), taken from the query set of \citet{clarke1995shortest} with notation adjusted for consistency with the modern Cottontail notation of Figure~\ref{fig:goth}.}
\label{fig:quebec}     
\end{figure}

%% file: results.tex
\begin{figure}[t]
\centering
\small
\begin{tabular}{l|r|p{0.62\linewidth}}
Run ID & NDCG@10 & Description \\
\hline
\texttt{uwc1} & \texttt{0.8199} & Multi-round LLM-based re-ranking of uwc0 \\
\texttt{uwc2} & \texttt{0.7971} & LLM-based re-ranking of track baseline \\
\texttt{monster} & \texttt{0.7933} & Reciprocal rank fusion (RRF) of uwc1 and uwc2 \\
\texttt{uwcCQAR} & \texttt{0.7169} & RankGPT re-ranking of uwCQA \\
\textbf{vole} & \textbf{\texttt{0.6863}} & Vole search agent (Section~\ref{sec:vole})  \\
\texttt{uwcCQR} & \texttt{0.6806} & RankGPT re-ranking of uwCQ \\
\texttt{uwcCQA} & \texttt{0.6765} & Dense retriever with LLM-based query expansion \\
\texttt{uwc0} & \texttt{0.6367} & RRF of 15 query expansion and ranker combinations\\
\texttt{uwcCQ} & \texttt{0.6177} & Dense retriever \\
\texttt{uwcBA} & \texttt{0.4581} & BM25F with pseudo-relevance feedback \\
\texttt{uwcBQ} & \texttt{0.3583} & BM25F with LLM-based query expansion \\
\end{tabular}
\caption{Performance of the Vole vs.\ retrieval runs submitted to TREC 2024 RAG Track by the first-author's team (WaterlooClarke\cite{monster}). Except for the Vole, NDCG@10 is based on the official track UMBRELA judgments, and all runs are fully judged to at least depth 10. For the Vole, which retrieved many unjudged documents, we filled holes with UMBRELA using the same model employed by the TREC 2024 RAG Track organizers (gpt-4o), uncovering 420 previously unjudged highly-relevant (grade~2) documents and 18 previously unjudged perfect (grade~3) documents. Without hole-filling the Vole achieves an NDCG@10 of 0.5340.}
\label{fig:results}     
\end{figure}